\documentclass[12pt]{article}

\usepackage[dvipsnames]{xcolor}
\usepackage{amsmath}
\usepackage{amssymb}
\usepackage{dsfont}
\usepackage[english]{babel}
\usepackage[utf8]{inputenc}
\usepackage{times}
\usepackage{graphics}
\usepackage{graphicx}
\usepackage[T1]{fontenc}
\usepackage[official,right]{eurosym}
\usepackage{url}
\usepackage{eso-pic}

\newtheorem{example}{Example}[section]

\title{What is legitimate decision support?} 
\author{Yves Meinard, Alexis Tsoukiàs, \\ CNRS-LAMSADE, PSL, Université Paris Dauphine}
\date{}

\begin{document}

\thispagestyle{empty}

\enlargethispage*{8cm}
 \vspace*{-38mm}

\AddToShipoutPictureBG*{\includegraphics[width=\paperwidth,height=\paperheight]{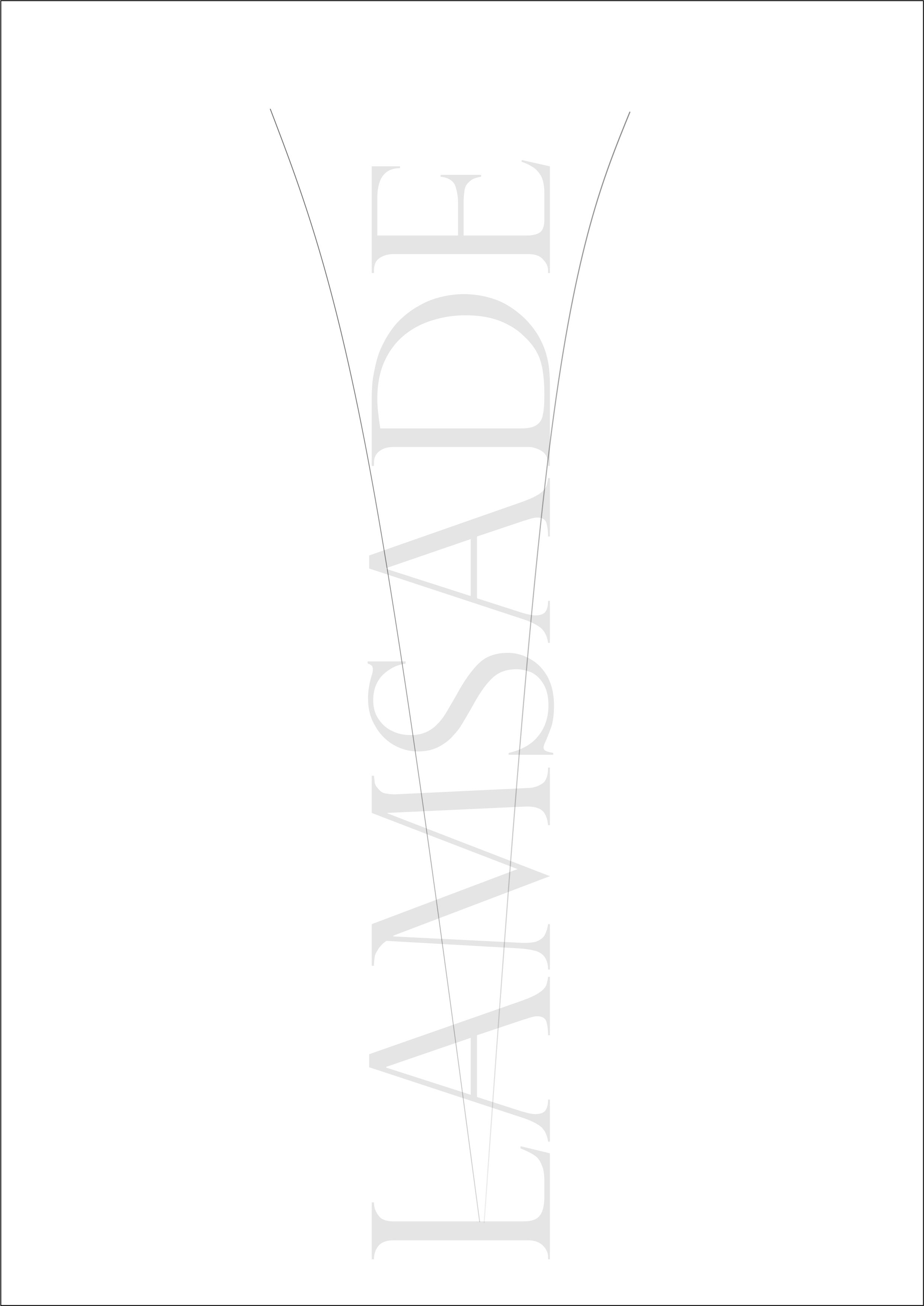}}

\begin{minipage}{24cm}
 \hspace*{-28mm}
\begin{picture}(500,700)\thicklines
 \put(60,670){\makebox(0,0){\scalebox{0.7}{\includegraphics{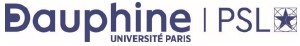}}}}
 \put(60,70){\makebox(0,0){\scalebox{0.3}{\includegraphics{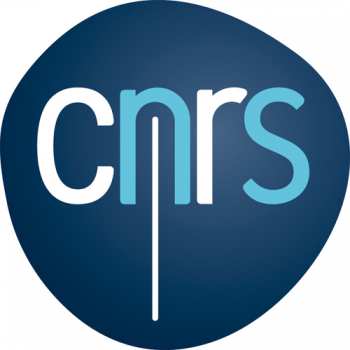}}}}
 \put(320,350){\makebox(0,0){\Huge{CAHIER DU \textcolor{BurntOrange}{LAMSADE}}}}
 \put(140,10){\textcolor{BurntOrange}{\line(0,1){680}}}
 \put(190,330){\line(1,0){263}}
 \put(320,310){\makebox(0,0){\Huge{\emph{401}}}}
 \put(320,290){\makebox(0,0){January 2022}}
 \put(320,210){\makebox(0,0){\Large{What is legitimate Decision Support?}}}
 \put(320,100){\makebox(0,0){\Large{Yves Meinard, Alexis Tsoukiàs}}}
 \put(320,670){\makebox(0,0){\Large{\emph{Laboratoire d'Analyse et Mod\'elisation}}}}
 \put(320,650){\makebox(0,0){\Large{\emph{de Syst\`emes pour l'Aide \`a la D\'ecision}}}}
 \put(320,630){\makebox(0,0){\Large{\emph{UMR 7243}}}}
\end{picture}
\end{minipage}

\newpage

\addtocounter{page}{-1}

\maketitle

\abstract{Decision support is the science and associated practice that consist in providing recommendations to decision makers facing problems, based on available theoretical knowledge and empirical data. Although this activity is often seen as being mainly concerned with solving mathematical problems and conceiving algorithms, it is essentially an empirical and socially framed activity, where interactions between clients and analysts, and between them and concerned third parties, play a crucial role. Since the 80s’, two concepts have structured the literature devoted to analysing this aspect of decision support: validity and legitimacy. Whereas validity is focused on the interactions between the client and the analyst, legitimacy refers to the broader picture:  the organisational context, the overall problem situation, the environment, culture, history. Despite its unmistakable importance, this concept has not received the attention it deserves in the literature in operational research and decision support. The present chapter aims at filling this gap. For that purpose, we review the literature in other disciplines (mainly philosophy and political science) that is demonstrably relevant to elaborate a concept of legitimacy useful in decision support contexts. Based on this review, we propose a general theory of legitimacy, adapted to decision support contexts, encompassing the relevant contributions we found in the literature. According to this general theory, a legitimate decision support intervention is one for which the decision support provider produces a justification that satisfies two conditions:  (i) it effectively convinces the decision support provider’s interlocutors (effectiveness condition) and (ii) it is organised around the active elicitation of as many and as diverse counterarguments as possible (truthfulness condition). Despite its conceptual simplicity, legitimacy, understood in this sense, is a very exacting requirement, opening ambitious research avenues that we delineate.}

\newpage

\section{Introduction}

Although the term "decision" might at first sight seem to refer a punctual event, in fact most decisions are made through a set of cognitive activities that the decision maker performs: a decision \emph{process}. Decision support is the science and associated practice that consist in providing recommendations to clients (possibly decision makers) facing problems, based on available theoretical knowledge and empirical data. Just like decisions or decision making, decision support is a process, rather than a punctual event. What do we do when, as decision analysts, we engage in a such processes \cite{Tsoukias07aor}? From an analyst's perspective, the answer is that we \emph{manipulate} information to provide recommendations. To formulate this idea, we purportedly use the ambiguous term "manipulate", because it conveniently conveys the idea that this task is double-edged. Indeed, depending on the context, "manipulate" can be either a neutral term, synonym for "compute" or handle", or be attached with negative connotations, and mean something more akin to "distort" or "falsify". When dealing with information in decision support processes, we are always in the grey zone between these two senses of "manipulate". On the one hand, we are mostly guided by a willingness to help the decision-maker, and in that sense we are not here to cheat or deceive her/him. But, on the other hand, when we work with information, using our data analyses technologies, our algorithms and theoretical and computational devices, we unavoidably make choices that are to some extent arbitrary, and about which the decision-maker does not have a say -- because we do not give her/him the opportunity, and/or because s/he does not have the required technical skills to make a cogent decision in this domain.

Usually, the information we "manipulate" in that sense consists in empirical observations, data collected in different forms and circumstances, as well as information about the subjectivity of the decision-maker, such as her/his values, beliefs and intentions. To these we may add norms, regulations, standards, which apply independently of the precise problem situation at hand, as well as culture, history, practices and habits that frame the space and time within which the decision support process is conducted. Although it is clear that the boundaries between these concepts might be blurry, for the sake of simplicity, we may categorise these different types of pieces of information in four categories:
\begin{itemize}
    \item objective data;
    \item preference statements;
    \item constraints that can be called "hard", in the sense that they are untouchable at the scale of the decision support process, such as laws and regulations, budgets constraints, and so on;
    \item cultural, or "soft" constraints, which are to some extent binding, but can nonetheless be somewhat slackened, such as habits and customs.
\end{itemize}

An extensive literature in Decision Sciences addresses the cognitive effort that gathering, computing and analyzing these information entails for clients and analysts developing decision support models (see \cite{greco2016multiple,winte2book86}), and associated biases plaguing decision aiding processes. An important outcome of these reflection is the development of ``user friendly'' methods (such as rule based decision support models; see \cite{12-MErgJFigSGre2005,SlowinskiGrecoMatarazzo2005}) and preference learning techniques (see \cite{DoumposZopounidisejor2011,FurnkranzHullermeier2010,MousseauPirlotejdp2015}).

Beyond these contributions, since the 80s, the bulk of academic discussions of the appropriateness of such "manipulations" have been mainly developed around two key concepts: validity and legitimacy.

Concerning the former, Landry (\cite{LandryMalouinOral83} see also \cite{LandryPascotBriolat83}) famously introduced four types of validity checks: conceptual, logical, experimental and operational. Seen from a decision support process perspective (focused on interactions between a client and an analyst), they can be regrouped in two categories:

1. To be valid, decision support should be meaningful for the analyst, in the sense that it should respect accepted axioms, theorems and properties. For example, the "manipulation" should respect meaning invariance (for more about the concept of meaningfulness in measurement theory see \cite{roberts79,Roberts85,Roberts94}).

2. To be valid, decision support should be meaningful for the client, in the sense that it should reply to her/his questionings, it should be useful in terms of advice on what to do (or not to do), it should be felt as owned by the client and usable within the decision process within which it has been requested.

However, as already noted by Landry himself in the 90s, although necessary, validity is not enough to ensure that the "manipulation" we produce and the recommendation that follows will be effectively used, applied and appreciated, and will have an impact in the real world. This is because validity refers to interactions between client and analyst, but ignores the larger picture: the organisational context, the overall problem situation, the environment, culture, history. Besides, more often than not, beyond the decision maker and stakeholders identified in the decision process, decisions also affect other stakeholders who can appreciate or not the decision, react to it by modifying their behavior, and \emph{in fine} influence how the whole decision process for which the decision support has been asked is conducted.

Although many analysts, especially in academic contexts, often pursue research interests when "manipulating" data, in decision support processes, we should keep in mind that such "manipulations" of information are not aimed at supporting the analyst, but the client. In other terms, whatever "manipulation" we do is going to be available for use by others, within contexts that we know and control only partially, producing consequences which most of the times will affect the client and the stakeholders involved in our recommendation. Hence, while many academic decision analysts tend to see their activity as mainly concerned with solving mathematical problems or conceiving algorithms, in fact supporting the decision making of others essentially is an empirical and socially defined activity \cite{Ackoff79a,Churchman67,moscarola84,rosen1book89,BR93}. This predicament means that, as analysts involved in decision support processes, we have to make sure that the "manipulations" we engage in, can make sense for the decision-maker in light of the context in which we support her/his decision making.

The concepts of validity and legitimacy are complementary in the sense that the latter encapsulates all the above-mentioned aspects of decision support interactions that the former, focused as it is on the interaction between client and analyst, fails to capture. When practicing decision support for their clients, analysts need not only check the validity of the information "manipulation" they perform, and the validity of the corresponding recommendation. They should also pay due attention to their legitimacy.

In this chapter, we propose and explain our vision of what the legitimacy requirement amounts to. This vision is aimed at clarifying debates on the desirable features of decision support processes, among other things by providing answers to the above questions.

For that purpose, we begin by showing that the issue of the legitimacy of decision support is both important and neglected in the current literature. We then proceed by reviewing the recent literature on legitimacy. Without claiming to be exhaustive, we will try to identify the main theoretical options exposed in the literature. Based on this review, we then propose a general theory of the legitimacy of decision support, designed to encompass the various visions presented in the preceding section, in such a way as to overcome their limitations and make the most of their strengths. This general theory is, to a large extent, based on preliminary discussions published in \cite{meinard_what_2017} and \cite{meinard_justifying_2020}. Equipped with this general theory, we will then be in a position to address a major, if relatively neglected question: the one of the challenges facing the quest for legitimacy in decision support contexts. Our aim in this section is to explore reasons why, despite all the major reasons  we have to take legitimacy serious (recalled above), in some cases the quest for legitimacy can prove extremely difficult, if not impossible to achieve. The exploration of theses hurdles on the path to legitimacy will enable us to identify a series of challenges for future research on means and approaches to construct the legitimacy of decision support.

\section{The legitimacy of decision support: an important but neglected topic}

The issue of the legitimacy of decision support is of unmistakable prominence when decision support activities are involved in policy making, policy design or policy evaluation \cite{jeanmougin_mismatch_2017,meinard_what_2017}. In such contexts, decision support is expected to improve or strengthen policies or parts thereof. The latter are activities that typically limit the liberties of some individuals and/or groups, and distribute financial or regulatory advantages among individuals and/or groups. This can, and often does, arouse questionings and disagreements. Besides, often enough, policies pertaining to different sectors compete with one another to capture public finances and public support (e.g. environmental policies vs. economic policies), and various such policies are involved in or criticized by the competing policy agendas of different political parties and/or candidates to elections in democracies. The corresponding debates around the well-foundedness of policies are often framed in the terminology of "legitimacy". The legitimacy of the decision support activities involved in the elaboration and implementation of the policies at issue is unavoidably raised as part of these debates.

The concept of legitimacy is, however, relevant to decision support well beyond political contexts \cite{ogien_politique_2021}. Even when decision support is deployed in private companies to address issues without any link with public policies, questions echoing the ones mentioned above unavoidably emerge. Indeed, the typical decisions for which decision support is requested in private firms, such as possible changes in strategy, reorganizations of the workforce or workflow or other organizational issues, typically have differential implications for various individuals within the organization: some individuals will gain prominence and/or responsibility, at the expanse of others, which can raise debates and disputes. Although the latter do not have, in private firms, the same importance as in political democratic arenas, still they can endanger the stability of the organization, which should accordingly pay attention to legitimacy.

This is all the more true when the decisions made in private firms have implications for public policies or, more generally, for the public. This is the case, for example, when a private firm decides to use a certain type of data or algorithm, which can involve the infringement of privacy or raise other stakes of public interest. Even beyond private organisations, as soon as issues of public interest can be involved in or impacted by decisions, the legitimacy of the decision maker and her/his decision and, consequentially, the legitimacy of the decision support s/he benefits from, are raised, even in the archetypal case of a single, self-standing decision maker. In the following we provide two short examples allowing to show the difference between constructing a valid model and providing a legitimate model for the decision process where the model is expected to be used.

\begin{example}{Example 1: Organisational Legitimization}
  The second author has been involved in the past in ``providing'' decision support to a large Italian company facing the problem of massive software acquisitions which needed to be framed by a general policy assisted by a rigorous evaluation model. The case study is reported in \cite{PaschettaTsoukias00}. When the whole study was completed, the final deliverable was almost dismissed by the General IT manager of the company because he considered it was too complicated for his staff to be effectively used. The project has been saved when the project manager revealed that the whole procedure was implemented on a spreadsheet (it has been largely used in the subsequent years). The reasoning of the IT manager was simple: if it runs on a spreadsheet it fits our organisational knowledge. Otherwise it is an academic exercise. On our side the reason for using the spreadsheet implementation was only rapid prototyping; actually the project manager was pushing for a very sophisticated (although user friendly) implementation. This is a typical case where a model certainly valid (for both the client and the analyst) risked to fail an organisational legitimacy check. It has been saved by chance.
\end{example}

\begin{example}{Example 2: Society legitimization}
  In the late 60s, early 70s the NYFD commissioned to the RAND corporation a large study concerning the location of the fire-fighters stations in order to improve the efficiency of the whole service and reduce the dramatic increase of casualties due to late intervention of fire brigades (see \cite{Walkeretal1979}). The project has been technically successful, but rose an extensive number of controversies both political and with the trade unions, resulting in reducing drastically the effectiveness of the suggested solutions (for a nice discussion see \cite{GreenKolesar2004}). This is a typical case where neglecting the social complexity of the problem at hand may produce ``valid'' models which turn to be socially unacceptable and thus not legitimated.
\end{example}

Despite this unmistakable importance, and associated major theoretical and practical implications, the issue of the legitimacy of decision support has not received the attention it deserves in the literature on decision science, operational research and management. \cite{LandryBanvilleOral96} in operational research, and \cite{suchman_managing_1995} in management sciences are notable, if now a bit old, exceptions. Despite their undeniable contribution (to be discussed below), they cannot compensate for the overall scarcity of discussions on this topic in this literature. By contrast, the literature on this topic is immense in a wide range of domains, from economics (e.g. \cite{vatn_environmental_2015}) to philosophy (e.g. \cite{habermas_faktizitat_1992}) and political sciences (e.g. \cite{backstrand_environmental_2010}). For lack of thorough, recent contributions to these debates in the specialized literature, the concrete meaning and implications of this large body of literature for the specific case of decision support are currently unclear.

Our aim here is to bridge this gap in the literature and, in so doing, hopefully, to bring our contribution to larger, interdisciplinary debates on the concept of legitimacy, from both theoretical and practical points of view.

\section{Visions of legitimacy}

\cite{LandryBanvilleOral96} notoriously emphasized the importance for models used in Operational Research practice to be legitimate, and they pointed the need to clearly make a difference between the validity of a model and its legitimacy. According to these authors, "legitimisation encompasses two complementary and often unconscious activities. The first one is a comparison of concrete actions, situations, or states of affairs with a set of abstract entities comprising values, norms or symbolic reference systems, which will be referred to as the 'code' henceforth. The second activity is a judgment as to the conformity of these concrete actions, situations, or states of affairs with the corresponding code." However, they do not clearly explain what they take this "code" to be.  Fortunately, a rich and profuse literature is available to clarify this issue, and overcome the limitations of \cite{LandryBanvilleOral96}'s seminal effort.

Discussions on legitimacy in the literature appear, at first sight, to be highly complex and dispersed, diversely focused as they can be on sources of legitimacy, criteria of legitimacy, means to ensure legitimacy, proofs of legitimacy, and so on. In this section, our aim is to draw a map of the main theories of legitimacy developed and used in the literature, to clarify this complex theoretical landscape.

The question of the legitimacy of a given decision support activity can be raised from two, complementary points of view: positive and normative. The positive approach asks an empirical question: what are the criteria that people use, \emph{as a matter of fact}, when deciding whether they take something to be legitimate or illegitimate? The normative question approach asks: what are the criteria that \emph{should} be used to decide if something is legitimate or not?

As scientists, we might think that the normative question is not for us, but for moralists or preachers, to answer, and that the positive question is the only one that can be addressed in a scientific context such as the one of decision support. But this would be mistake, for two associated reasons.

First, as famously explained by \cite{putnam_collapse_2004}, the frontier between normative and positive is always blurred, since many issues, theories or approaches that are typically seen as entirely positive in fact have a normative anchorage, and most issues that are typically seen as entirely normative are based on, or influenced by, positive data. Confining ourselves to the positive question is accordingly impossible in practice. In concrete terms, this impossibility stems from the fact that, when designing a scientific project to answer the empirical question above and when analysing the data obtained, we unavoidably take stances on issues that pertain to the normative approach. This is the case, for example, when making decisions on how questions will be formulated to survey individuals, or on how behaviour will be monitored and interpreted.

The second, related reason is that, most of the time, there is no such thing as a "fact of the matter" when it comes to what people take to be legitimate or illegitimate. People might fail to have an opinion on what they take to be legitimate or not. They might start asking themselves the question and forming an opinion upon our asking them. They might change their mind if we give them pieces of information, perhaps even if this information is irrelevant. On issues such as legitimacy, the picture according to which people always already have a well-formed, stable vision, independent of the scientist and the scientific protocol that strive to capture this independent "fact of the matter" is accordingly untenable.

The crude vision according to which normative questions are for preachers or moralists to address is therefore entirely irrelevant when issues such as those surrounding legitimacy are raised. Normative philosophy is, for that matter, to a large extent devoted to address normative questions in a rational way, rather than through preaching. In contemporary normative philosophy, Rawls (\cite{Rawls1971}) and Habermas (\cite{habermas_theorie_1981}) are the most prominent authors who have championed this rationalist approach to normative philosophy, in the wake of Kant's philosophy of practical reason. Such philosophical approaches to normative questions do not evacuate positive questions: they strive to take advantage of studies of positive questions to enrich normative reflection, and conceive of the latter as relevant to improve the way positive questions are addressed. In the remainder of this chapter, we will endorse a similar approach. We will take into account both normative and positive approaches to legitimacy, and we will strive to use both as complementary approaches liable to enrich one another.

Beyond the normative/positive dichotomy of points of view, visions of legitimacy in the literature are classically divided into two broad categories: theories of output legitimacy, also called substantive theories \cite{vatn_environmental_2015}, and theories of input legitimacy, also called procedural theories \cite{backstrand_environmental_2010}. Substantive theories claim that the legitimacy of a policy or decision depends on the state of affairs that it brings about. By contrast, procedural theories hold that the legitimacy of a decision is determined by the decision making process through which the decision was made. A toy example of discordance between substantive and procedural visions of legitimacy can be given by the following scenario: imagine that, through a democratic decision making process such as a majority vote, a minority group in society is denied some basic rights, such as access to education and health insurance. A plausible procedural theory of legitimacy might claim that the decision is legitimate, because it was made through a legitimate process (majority vote). A plausible, dissenting substantive theory might claim that such a policy that end up arbitrarily depriving some people from some basic rights cannot be legitimate, because the state of affairs in which a minority is oppressed is illegitimate.

This substantive/procedural dichotomy is useful to clarify some debates on legitimacy, since numerous theories can easily be classified along the lines of this dichotomy. Among prominent visions of legitimacy than can be classified in this way, take for a example a vision according to which the essence of legitimacy is due process or legality. In this vision, a decision is legitimate if it rigorously abides by all the relevant regulatory rules. This first vision clearly falls in the procedural category. Similarly, a vision according to which the effective participation of citizens is the crux of legitimacy is another example of a plausible procedural theory. By contrast, a theory claiming that a policy is legitimate if it ensures that all the people affected see their welfare increased, falls in the substantive category. The same goes for theories of so-called "higher goods", such as the one championed by \cite{taylor_sources_1989}. Among theories clearly falling into one or the other category, a diversity of concrete criteria through which legitimacy is ascertained can then emerge: procedural theories can champion criteria of fairness, impartiality, responsibility, while substantive theories will use criteria of equality, efficiency, or effectiveness.

For all its usefulness for clarification purposes, the substantive/procedural dichotomy has, however, its limits. Indeed, numerous theories of legitimacy mix procedural and substantive aspects. This intimate mix of substantive and procedural aspects can even be found within some basic concepts that can hardly be avoided in discussions on legitimacy. This is the case, for example, of the concept of right. On the one hand, the picture of who enjoys a given right and who is deprived from it in a given population is, in a sense, a state of affair. In that sense, a theory that would hold that this right is the crux of legitimacy would be called substantive (this is what we have done in our example above). But, on the other hand, a right is procedural, in the sense that a right specifies what people who enjoy it or are deprived from it can do. This is particularly evident when the right at issue is a right to vote or to partake in a decision, but this is true of rights in general. The theory holding that a right is the crux of legitimacy is, in that sense, also procedural. The same logic applies at least to some values, as illustrated, for example, by \cite{brettschneider_democratic_2007}'s theory of democracy. According to this author, the essence of democratic legitimacy is a set of "core values" than can materialize in both procedures (hence the procedural aspect of his theory, which refers to voting and parliamentary procedures) and substantive judgements (made by judicial courts, such as the Supreme Court in the United States).

Another weakness of the substantive/procedural dichotomy is revealed by "epistemic" theories \cite{estlund_democratic_2009}. These theories focus on procedures to decide if a decision is legitimate or not. But they do not take procedures to be the core of legitimacy, which they locate in substantive features. They focus on procedures because they take them to be the most reliable means we have to make sure that the substantive features of interest are and/or will be brought about.


\section{The legitimacy of decision support: a general theory}

At this stage, we hence see that, although there is a large diversity of visions of legitimacy, a series of concepts (the normative/positive and substantive/procedural dichotomies, the notion of epistemic approaches, etc.) can be put to use to clarify the complex picture that this large diversity of visions draws. We have also seen that these various concepts have their limits. But they can be used as complementary tools, whose limited relevance should be assessed on a case-by-case basis when using them, to clarify debates on legitimacy.

Now that we have this complex landscape and a set of conceptual tools to navigate this landscape, the question should be asked: is it possible to elaborate a unique, central theory of legitimacy, possibly useful to think through concrete issues, such as the ones associated with decision support activities?

We argue that the literature in normative philosophy on deliberative democracy \cite{chappell_deliberative_2012,dryzek_deliberative_2002,habermas_faktizitat_1992,rawls_political_2005} provides the key to overcome the diversity of visions of legitimacy, so has to develop a unique, encompassing theory. Just like our overview of legitimacy witnesses a diversity of theories, theories of deliberative democracy are concerned with situations in which a diversity of ethical views co-exist and are championed by a diversity of people and/or groups composing a society. In this context primarily characterized by \emph{pluralism}, theories of deliberative democracy are concerned to identify means to make collectively acceptable decisions, without hoping to identify decisions that will perfectly match any one of the diverse points of view that are concerned. The key concept through which theories of deliberative democracy claim to escape chaos is \emph{justification}. According to theories of deliberative democracy, decisions can be collectively made in a pluralist setting if they can be justified to all the diversity of concerned actors or groups.

This idea raises numerous questions that fall beyond our scope in this paper, such as: how can one be sure that it will be possible to justify a given policy to all those concerned? How should we identify who are those people that are called "concerned"? Etc. We leave aside these questions here because our point is not to champion the theory of deliberative democracy, but to assess if the reference to justification, which is used by this theory to address pluralism, can be used in our case to address the diversity of visions of legitimacy.

Beyond the similitude in context (deliberative democracy faces a diversity of ethical views, and we face a diversity of visions of legitimacy), the idea to use the same concept of justification stems from the way deliberative democracy uses this concept, which appears relevant to our case as well. Indeed, the crux of the usage of the concept of justification in deliberative democracy consists in taking the various ethical views composing pluralism as a reservoir of \emph{building blocks for candidate justifications}. Seen from theses lenses, any given ethical theory contains, or can lead to the formulation of, justifications for some decisions but not for others. These justifications will, typically, be accepted by people endorsing this ethical view, but probably not by people endorsing other ethical views. In a deliberative dynamics, such disagreements should lead to the formulation of new justifications, less directly anchored in any given ethical view, and therefore liable to enable agreement among a diversity of people endorsing different ethical views. This approach to diversity and pluralism suggests that, in our case of a diversity of visions of legitimacy, the various visions can also be seen as reservoirs of building blocks for justifications.

In this general approach, legitimacy is a matter of justification, and the various visions of legitimacy found in the literature are partial justifications that can complement and enrich one another in various situations, as required by the context. Various applications of this or that substantive theory, or this or that procedural theory, understood in a normative or a positive interpretation, should hence be seen as elements that can be combined to produce justifications. Some combinations will prove incoherent, others irrelevant, beside the point, unnecessarily intricate, and so on. But some combinations might constitute convenient justifications in some case.

However, this usage of the concept of justification creates, at this stage, an important problem, heralded by the fact that, in the last sentence, we have had to add an adjective ("convenient") to qualify justifications. This need to qualify justifications stems from the fact that the term "justification" is, as it stands, ambiguous. Indeed, what, precisely, is a justification? A basic idea conveyed by this term, which we posit is shared by all the users of the term, is that a justification is an argumentative discourse. But our usage of the concept of justification in our general theory cannot be limited to this basic idea. Indeed, it would not make sense for us to claim that articulating an argumentative discourse that would be incoherent or nonsensical or beside the point is enough to yield legitimacy. Hence the need to \emph{qualify} the kinds of argumentative discourses that are relevant for our purposes.

We argue that two complementary qualifications are needed to equip ourselves with a relevant notion of justification: the first has to do with \emph{effectiveness}, the other with \emph{truthfulness}. To explain the meaning of these two qualifications, let us focus on our core setting of interest: decision support activities involving decision support providers (typically, decision analysts or experts), decision makers and concerned stakeholders.

If the decision support provider is concerned to entrench the legitimacy of her/his intervention, according to our general theory, s/he will elaborate and voice a justification. But if no one understands her/his argument, or if it fails to convince anyone, it is clear enough that her/his justification will have failed to yield legitimacy. We therefore need to add an \emph{effectiveness} requirement to the meaning we give to the concept of justification within our theory of legitimacy: what is needed, as a matter of justification, is an argumentative discourse \emph{that manages to convince the relevant public}.

But this reference to \emph{effectiveness} immediately raises two problems.

The first problem is just as ancient as philosophical reflections on speech and its ambivalent relation to rationality \cite{cassin_rhetorique_2015}. This problem is that, if we focus uniquely on \emph{effectiveness}, we will end-up with a wholly manipulative concept of justification (in the negatively connoted sense of the term). This would lead to an absurd approach in which the more manipulative (still in the negatively connoted sense of the term) the decision support provider is, the more legitimate her/his intervention is (this echoes \cite{LandryBanvilleOral96}'s claim that "legitimisation cannot be mystification" and that we "should not confuse manipulation and legitimisation"). Therefore, we need another qualification, designed to ensure that \emph{effectiveness} does not stem from mystification, but from rational persuasion \cite{perelman_traite_1958}. Because the point of this qualification is to ensure that effectiveness does not reflect mystification, but rather a faithful account of relevant facts and theories, let us talk about a \emph{truthfulness} qualification.

The second problem, which can be called the problem of "the targets of justification," is that, if we accept to abide by an effectiveness requirement, the question unavoidable arises: effective \emph{for whom}? Should the justification be convincing for the decision maker and only for her/him? Should it also include those actors who are tightly involved in the decision making process, such as members of a steering committee monitoring the process when one such committee exists? Should the circle of interlocutors to be convinced include all concerned stakeholders, or all the people that see themselves as potentially impacted by the decision to be made, or all the people that can take a stance on the issue even though they can not be directly impacted? The theory and practice of deliberative democracy and participation face notorious difficulties to answer such questions. Participatory practices usually informally choose the stakeholders that are asked to participate, and despite academic calls to formalize stakeholders' recruitment \cite{nabatchi_putting_2012}, there is currently no largely accepted technology available for that purpose. This lack of practical solutions reflects a theoretical difficulty, which is unmistakable in the main theoretical works on deliberative democracy. \cite{rawls_political_2005}'s idea to solve this problem was that justifications should be acceptable to all "reasonable" citizens, and he claimed that the precise content of this requirement should be clarified by reasonable citizens themselves. However, as \cite{estlund_insularity_1998} has shown (and as anyone should have expected), this purported solution does not work, since there is an "impervious" plurality of groups that might call themselves "reasonable". As opposed to Rawls's (untenable) refusal to clarify what "reasonable" means, \cite{estlund_democratic_2009} claims that philosophers of deliberative democracy should acknowledge that a "true" theory of who is reasonable and who is not is needed. But Estlund does not explain how this "truth" is to be discovered. Rawls's and Estlund's theoretical stances, which are the two options developed in the theoretical literature, therefore fail to solve the problem of the targets of justification.

We argue that these two problems can be solved by designing a truthfulness qualification fit for purpose.

Identifying means to ensure that a justification is truthful rather than manipulative is a notoriously difficult question. Here, we propose to take advantage of \cite{meinard_justifying_2020}'s approach, introduced in the context of a reflection on the justification of norms underlying decision support, to solve this problem. This approach proposes that, when developing a justification, one should actively seek as many counter-arguments as possible, including by soliciting the interventions of outsiders and people marginalized from the decision support process, and then enrich one's justification by defending it against all these counter-arguments. The underlying idea is that mystifying arguments typically stress convenient aspects of the matter, while silencing inconvenient aspects. A powerful counter-manipulative tactic is therefore to track aspects that presumably mystifying discourses tend to silence. By organizing one's justifications around an active search for counter-arguments, one therefore puts oneself in a position in which being mystifying is by design extremely difficult. By the same token, the "target of the justification" problem is solved. Indeed, if the search for counter-arguments is thorough enough, and if the defense against all these counter-arguments is effective, the justification will by definition be convincing \emph{to all}.

At this stage, a natural rejoinder might be to claim that the idea of "actively seeking as many counter-arguments as possible" is exceedingly vague and easily manipulable: if the decision support provider concerned to produce a justification takes, say, five minutes to seek counter-arguments, is it enough? And how "active" should s/he be? The notion of "active search" might appear much too indeterminate. But, as \cite{meinard_justifying_2020} argue, this indeterminacy would be a serious flaw of the theory only if the latter had the pretension to achieve an "absolute" justification, taking into account all the possible counter-arguments, from absolutely all sides. Achieving such an "absolute" justification is, in any case, impossible, since the universe of counter-arguments is infinite, and there even exists an infinity of counter-arguments that have not yet been discovered. As opposed to this unreachable "absolute justification," the truly worthwhile pursuit is the search for the best locally achievable justification, while keeping in mind that justifications are always provisional: new counter-arguments can emerge, and ruin an hitherto convenient justification, and alternative decision support interventions can be launched, and be supported by justifications that overcome the former one.

To sum up, our general theory of the legitimacy of decision support interventions, which we claim encompasses all the other theories reviewed above, is the following: \textit{``A legitimate decision support intervention is one for which the decision support provider (or, for that matter, anyone else), produces an unavoidably provisional justification that satisfies two conditions: (i) it effectively convinces the decision support provider's interlocutors (effectiveness condition) and (ii) it is organised around the active elicitation of as many and as diverse counterarguments as possible (truthfulness condition)''}.

In the following we provide a short final example in order to show how our theory of legitimacy would apply in a recent real world case study.

\begin{example}{Example 3: The legitimacy of using predictive justice tools}
In the past few years, there were extensive discussions about the use, abuse and misuse of predictive justice devices. The best known controversy is the ``COMPASS'' case\footnote{ \url{https://www.propublica.org/article/machine-bias-risk-assessments\\-in-criminal-sentencing}} (for a nice discussion, see \cite{AbuElyounes2020}): it concerns the fact that, behind a device computing a ``score'' which is used in order to assist a decision maker (a judge in this case) in making a decision, there are ``hidden'' hypotheses and assumptions. In that precise case, these hidden assumptions refer to manipulations that can be considered to be ``racial discrimination'' when the software computes the score for people from different racial origins.

A standard way to analyze such a case consists in striving to show that, because the tool manipulates data on racial origin in a certain way, it is unfair. However, fairness is a concept with several different formal definitions and it turns out that many such definitions are incompatible. Fairness, as a general and vague concept, can be used both to defend the use of data on racial origins and to dismiss it as discriminatory. This standard way to discuss the case is therefore inconclusive, because the concept that is supposed to play the key role in criticizing the tool, turns out to be ambiguous.

 This case easily lends itself to an alternative approach, along the lines suggested by our theory of legitimacy. Instead of focusing on fairness, our approach suggests that the problem with the tool is not that it is unfair, but that the vision of fairness it presupposes has been imposed without discussion, in an opaque way, without any justification or explanation whatsoever. From the point of view of our theory, the tool is hence illegitimate stems because relevant criticisms that can be raised against it are swept under the carpet. Implementing our approach to legitimacy here would consist in actively searching for such criticisms (truthfulness conditions), and setting out to convince decision makers, concerned stakeholders and other interlocutors that the decisions made, and the associated authorizations and bans, are meaningful and relevant (effectiveness condition) -- even if this means, in the end, that some of the procedural features currently structuring the process, or the final decision itself, will have to be adjusted to become more legitimate.
\end{example}

\section{Hurdles on the road to legitimacy}

The general theory of legitimacy presented in the former section is simple enough in its formulation. Its basic components (the production of argumentative discourses constituting justifications of decision support interventions, the test of the extent to which these justifications are convincing, and the active search for counterarguments), are activities in which anyone can engage, more or less successfully. Many practicing decision support providers certainly already engage in these activities, informally and to some limited extent, in their everyday decision support interventions. However, to go beyond such informal, unchecked practices, there is a need to organize this legitimization endeavor in a systematic and formal way. As we will show in this section, despite the \emph{prima facie} simplicity of the activity that consists in producing the kind of justifications structuring the above general theory of legitimacy, its formal systematization is deeply challenging. In this section, we will present what we take to be the two most important hurdles complicating the accomplishment of the legitimization task. These hurdles represent as many avenues for future research on the legitimacy of decision support interventions.

The first, and most evident, challenge, is to elaborate operational methods and tools to support the various steps of the production of justifications. This involves elaborating and deploying technologies to search for counter-arguments, including the search for and elicitation of neglected and/or marginalized sources of counter-arguments. Because marginalized sources of arguments typically use means of expression that are different from the mainstream ones, integrating them in argumentative discourses will also generate difficulties. Besides, in cases in which relevant counterarguments will be numerous and complex (which seems bound to be the general case, except for the most trivial applications), a risk will be that argumentative discourses including all those counter-arguments could become too complex, long and convoluted to be understandable. There is hence a real challenge to construct readable and accessible argumentative architectures based on such a complex and profuse material. Informal \cite{habermas_theorie_1981,perelman_traite_1958} and formal \cite{amgoud_using_2009,besnard_elements_2008,dung_acceptability_1995} approaches to argumentation theory will certainly prove useful to address these operational challenges. However, as explained by \cite{cailloux_formal_2020}, as they stand, these approaches are ill-equipped to address the empirical dimensions of these challenges. This is because this literature does not explore how decision support providers can organize their interactions with decision makers and other interlocutors, so as to assess how convincing various arguments are, without mystifying them (see \cite{cailloux_formal_2020} and \cite{cailloux_formal_2020,meinard_justifying_2020}, for a deeper exploration of this first research frontier).

A second, perhaps even more difficult hurdle on the road to legitimacy, refers to what one might call "mediation" in justifications of decision support interventions: that is, the intervention of third parties in interactions between producers and receptors of justifications. Two kinds of mediation play a prominent role in many decision support interactions:

\begin{itemize}
    \item \emph{Representation.} Decision support interactions only rarely involve the direct participation of all the actors potentially concerned by the decision at issue. Decisions are rather typically made in small circles, including the formal decision maker(s) and the decision aid provider. Over the last decades, the inclusion of stakeholders in these circles has been increasingly championed (see e.g. \cite{runge_structured_2020}), and the participation of stakeholders in decision making is now commonplace, through various organisational devices, such as steering committees. In such settings, vasts groups of stakeholders are typically represented by a tiny sample of "representatives", including elected representatives, trade-unionists, agents working for institutions allegedly representing various issues of public interest, or simply individuals who see themselves and are seen by others as "typical" of a larger group of concerned people.
    \item \emph{The "nesting" of decision support interactions.} In typical decision support interactions, decision support providers and experts with whom they interact often take advantage of, use or refer to various kinds of outcomes of antecedent or parallel decision support interactions: reports produced when trying to solve similar problems in other contexts, tools such as software or databases constructed in other contexts, methodological reports elaborated on the basis of series of similar missions, scientific publications, etc. In so doing, decision support providers take the role of decision makers supported by other decision support providers, who are themselves, for the same reason, decision makers benefiting from antecedent or parallel decision support interactions. These various decision support interactions can take different forms and, typically, the higher up we climb the hierarchy, the less interactive the "interaction" will be. For example, as authors of this paper, we are decision makers supported by, among others, Pythagoras and Aristotle, as great figures in our intellectual formation. But our "interaction" with them is much less interactive than the one we have with clients for whom we work as decision support providers. Anyways, through references and the usage of tools, decision support interactions are all nested in an infinite series of other, more or less clearly defined and formalized, decision support interactions.
\end{itemize}

Both aspects of mediation substantially complicate the task to produce legitimizing justifications:
\begin{itemize}
    \item Representation raises the questions:  if we manage to produce a justification that truthfully convinces representatives, can we admit without further ado that it is enough?  Should not we rather strive to convince those people who are supposed to be represented by the representatives?  What if we manage to convince representatives but not represented persons, or the other way round?
    \item The nesting of decision support interactions raises the questions: how far should we go when we decide on the aspects of the decision support interaction that we should justify? Evidently enough, we cannot set ourselves the requirement to justify each and every aspect of the decision support interaction, since this would mean, for example, that we would have to justify all the aspects of the foundations of the mathematical theories on which our theories and tools are based. But then, how are we to make a choice between the various aspects that could be justified?
\end{itemize}

At this stage, we do not claim to be able to answer these difficult but unavoidable questions. They constitute major agendas for further research.

\section{Conclusions}

Supporting the decision activities of some client (potentially a decision maker) can certainly be characterised by the use of formal and abstract models. However, pragmatically it is more complex than a simple application of such models. Although the topic of model validity has been discussed in the literature and can be based on some formal requirements (such as meaningfulness), there is a problem of model and more generally of decision support legitimacy.

In this paper we show that, with the notable exception of some seminal contributions, this topic is essentially neglected and barely developed. Renewing a tradition of discussions which used to animate the meetings of the EURO MCDA Working Group, this paper aims at suggesting a new perspective on decision support legitimacy.

We have proposed a general theory of the legitimacy of decision support processes, and used examples to illustrate the importance of the topic and the application of our theory. At the end of the day supporting our clients within their decision processes consists in convincing: \\
 - ourselves, that we appropriately used our models and methods; \\
 - our clients, that what we suggest and advise makes sense for them; \\
 - any other potential stakeholder, about the potential impact of this advice.

Our contribution is far from being exhaustive. Our topic has multiple theoretical and practical extensions and research pathways that could not be explored here. We hope that our broad community will pursue our effort by exploring these other aspects of the topic. Among the questions that researchers should address in this future effort, the most prominent ones are perhaps:
\begin{itemize}
\item which are or can be considered to be legitimate sources of information?
\item what does it mean to perform a legitimate information manipulation?
\item who is expected to release a ``patent of legitimacy'' within a decision support process?
\end{itemize}

\bibliographystyle{plain}
\bibliography{legitimacy}

\end{document}